\documentclass[a4paper,useAMS,usenatbib]{mn2e}
\usepackage[final]{graphics}
\usepackage{epsfig}
\usepackage{amssymb}

\newcommand{\hel}[2] {He\,{\sc #1}~$\lambda$#2}
\newcommand{\kms}{\mbox{$\mathrm{km~s^{-1}}$}}

\newcommand{\Line}[3]{\ion{#1}{#2}~$\lambda$#3}
\newcommand{\Lines}[3]{\ion{#1}{#2}~$\lambda\lambda$#3}
\newcommand{\ion}[2] {#1\,{\sc #2}}
\newcommand{\Ha}{\mbox{${\mathrm H\alpha}$}}
\newcommand{\Hb}{\mbox{${\mathrm H\beta}$}}

\title[Intermittent accretion in BB Dor]{The fight for accretion: discovery of intermittent mass transfer in BB Doradus in the low state}
\author[P. Rodr\'{i}guez-Gil et al.]{P. Rodr\'{i}guez-Gil$^{1,2,3}$\thanks{E-mail: prguez@iac.es (PRG)}, L. Schmidtobreick$^{4}$, K. S. Long$^{5}$\thanks{Visiting astronomer, Cerro Tololo Inter-American Observatory, National Optical Astronomy Observatory, which are operated by the Association of Universities for Research in Astronomy, under contract with the National Science Foundation.}, B. T. G\"ansicke$^{6}$,
\newauthor
M. A. P. Torres$^{7,8}$, M. M. Rubio-D\'{i}ez$^9$, M. Santander-Garc\'{i}a$^{10}$\\
$^{1}$Departamento de Astrof\'\i sica, Universidad de La Laguna, La Laguna, E-38204, Santa Cruz de Tenerife, Spain\\
$^{2}$Instituto de Astrof\'\i sica de Canarias, V\'\i a L\'actea, s/n, La Laguna, E-38205, Santa Cruz de Tenerife, Spain\\
$^{3}$Isaac Newton Group of Telescopes, Apartado de Correos 321, Santa Cruz de La Palma, E-38700, Spain\\
$^{4}$European Southern Observatory, Casilla 19001, Santiago de Chile, Chile\\
$^{5}$Space Telescope Science Institute, 3700 San Martin Drive, Baltimore, MD 21218, USA\\
$^{6}$Department of Physics, University of Warwick, Coventry CV4 7AL, UK\\
$^{7}$SRON, Netherlands Institute for Space Research, Sorbonnelaan 2, 3584 CA, Utrecht, the Netherlands\\
$^{8}$Harvard--Smithsonian Center for Astrophysics, 60 Garden Street, Cambridge, MA 02138, USA\\
$^{9}$Centro de Astrobiolog\'{i}a (CSIC-INTA), Ctra. de Torrej\'on a Ajalvir km-4, Torrej\'on de Ardoz, E-28850, Madrid, Spain\\
$^{10}$Observatorio Astron\'omico Nacional, Apdo. de Correos 112, Alcal\'a de Henares, E-28803, Madrid, Spain\\}
\begin{document}
\date{Accepted 2012. Received 2012}
\pagerange{\pageref{firstpage}--\pageref{lastpage}} \pubyear{2012}
\maketitle
\label{firstpage}

\begin{abstract}

Our long-term photometric monitoring of southern nova-like cataclysmic variables with the 1.3-m SMARTS telescope found BB Doradus fading from $V \sim 14.3$ towards a deep low state at $V \sim 19.3$ in April 2008. Here we present time-resolved optical spectroscopy of BB Dor in this faint state in 2009. The optical spectrum in quiescence is a composite of a hot white dwarf with $T_\mathrm{eff} = 30000 \pm 5000$\,K and a M3--4 secondary star with narrow emission lines (mainly of the Balmer series and \ion{He}{i}) superposed. We associate these narrow profiles with an origin on the donor star. Analysis of the radial velocity curve of the \Ha~emission from the donor star allowed the measurement of an orbital period of $0.154095 \pm 0.000003$ d ($3.69828 \pm 0.00007$ h), different from all previous estimates.  We detected episodic accretion events which veiled the spectra of both stars and radically changed the line profiles within a timescale of tens of minutes. This shows that accretion is not completely quenched in the low state. During these accretion episodes the line wings are stronger and their radial velocity curve is delayed by $\sim 0.2$ cycle, similar to that observed in SW Sex and AM Her stars in the high state, with respect to the motion of the white dwarf. Two scenarios are proposed to explain the extra emission: impact of the material on the outer edge of a cold, remnant accretion disc, or the combined action of a moderately magnetic white dwarf ($B_1 \lesssim 5$ MG) and the magnetic activity of the donor star.   
\end{abstract}


\section{Introduction} \label{intro}

The highest mass transfer rates among cataclysmic variables (CVs) are found in the nova-like class. Nova-likes are non- or weakly-magnetic CVs typically found in a state of high accretion rate with a hot, steady state disc. As a consequence of the high accretion rate, the accretion disc in these systems is usually too bright to allow detection of the white dwarf (WD) and the donor star, thus preventing dynamical studies aimed at measuring fundamental binary parameters from being done. Fortunately, nova-likes are occasionally caught in faint states, when their brightness can drop by $\sim$3--5 mag. Their observation during these low states (also known as VY Scl states) is the only practical opportunity to study the WDs and donor stars in these binary systems, especially the magnetic activity of the fast rotating donors, their solar-like activity cycles and the interplay with the WD magnetic field or a remnant accretion disc if present.

Of particular interest are the low states of nova-like CVs which gather in the 3--4 h orbital period regime. They lie just at the upper boundary of the period gap, where donor stars are predicted to become fully convective and orbital angular momentum loss via magnetic wind braking is expected to greatly diminish or even cease \citep[e.g.][]{rappaportetal83-1}. Furthermore, at least 50 per cent of all CVs in that period range belong to the SW Sex class \citep{rodriguez-giletal07-1}, characterized by having large average mass transfer rates leading to very hot WDs \citep{townsley+bildsten03-1,araujo-betancoretal05-2,townsley+gaensicke09-1}.

While it is generally accepted that mass transfer from the donor star to the WD through the inner Lagrangian (L1) point temporarily stops or is greatly decreased during a low state, the exact mechanism is not known. \cite{livio+pringle94-1} have proposed that accumulation of starspots in the vicinity of the L1 point may inhibit Roche-lobe overflow. With the main mass transfer channel suppressed, the luminous accretion disc wouldn't be fed as efficiently as in the high state. The survival of the accretion disc during the low state is still a matter of debate. \cite{leachetal99-1} and \cite{hameury+lasota02-1} have invoked a truncated accretion disc to explain the absence of dwarf nova outbursts during the decay to the low state. The former authors propose irradiation of the inner accretion disc by the hot WD as a means of keeping the internal parts of the disc in a hot, viscous state, while the latter suggest the total absence of the inner parts due to the action of a magnetic WD. Both scenarios suggest that the accretion disc conserves most of the mass it contained in the high state. However, observations show no contribution of accretion discs to the optical light of the binary system in the low state. In this regard, \cite{gaensickeetal99-1} studied the low state of TT Ari and found that any remnant accretion disc should be optically thin up to a distance from the WD of 12 WD radii and have an effective temperature less than 3000\,K to be consistent with the low-state optical spectra. Similarly, the modelling of the spectra of MV Lyr in the low state reported in \cite{linnelletal05-1} did not require an accretion disc. In case a disc existed in MV Lyr during the low state it would be cooler than 2500\,K to avoid discrepancy with their optical spectra.
 
Under low state conditions, only gas expelled from the donor star due to, for example, magnetic activity or the stellar wind would then be available for accretion \citep[e.g.][]{hessmanetal00-1}. The presence of a WD with an intense magnetic field ($B_1 \gtrsim 1$ MG) interacting with that of the donor star would lead to an alternative accretion channel. In fact, entanglement of the magnetic fields of both stars has been proposed to explain the line emission patterns observed in AM Her, the prototypical magnetic CV, during low states \citep[see][and references therein]{kafkaetal08-1}.

\begin{figure}
\includegraphics[width=8.8cm]{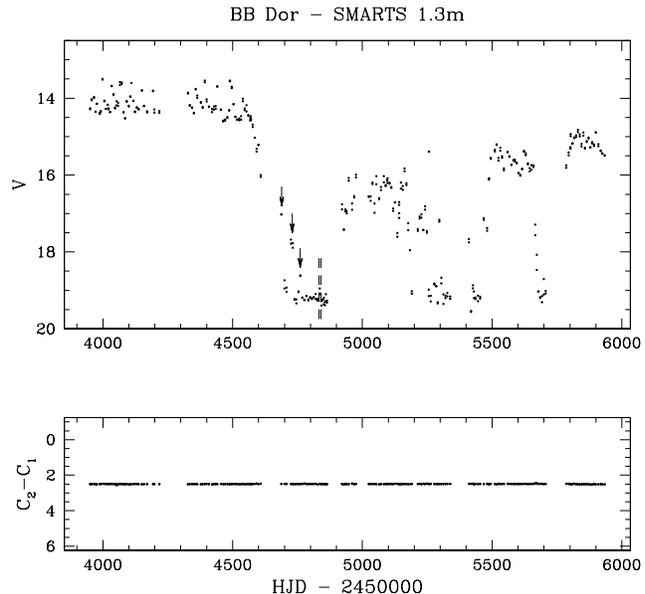}
\caption{{\em Top panel}: Long-term $V$-band light curve of BB Dor showing 5.43 years of photometric coverage with the 1.3-m SMARTS telescope on Cerro Tololo.  The two vertical dashed lines mark the week of NTT spectroscopic observations on 2009 Jan 1--7. The arrows mark the brightenings observed during the decline to quiescence. {\em Bottom panel}: differential light curve of the comparison star $\mathrm{C_1}$ and the check star $\mathrm{C_2}$.}
\label{fig_longtermphot}
\end{figure}

It is known that about half the nova-likes which have experienced a low state (collectively named as the VY Sculptoris stars) are SW Sex stars \citep{rodriguez-giletal07-1}. In addition, there is increasing evidence of the presence of magnetic WDs in this class of CV provided by the detection of variable circular polarization \citep{rodriguez-giletal01-1,rodriguez-giletal02-1,rodriguez-giletal09-1}, variable X-ray emission \citep{baskilletal05-1}, and emission-line flaring \citep[][and references therein]{rodriguez-giletal07-1}. The potential similarities between the behaviour in the low state of the strongly magnetic polar CVs and the SW Sex stars led us to start a long-term monitoring campaign to search for low states in the latter. Once in the low state with a negligible contribution of an accretion disc, the component stars and any other accretion processes are open to study.   

In this paper we present the results of such a study for the nova-like CV BB Doradus. BB Dor (= EC 05287$-$5857) was first identified as a CV in the Edinburgh--Cape Blue Object Survey \citep{stobieetal87-1}. \cite{chenetal01-1} classified BB Dor as a low-inclination VY Sculptoris star based on long-term photometry, which showed the system undergoing distinct low and high brightness states ($\Delta B \sim 2.5$) as well as rapid variability at times. No coherent orbital modulation was found in their photometric light curves. However, a very uncertain orbital period of $P= 0.107 \pm 0.007$ d was estimated from the analysis of radial velocities. \cite{pattersonetal05-3}, as part of a long-term photometric campaign to search for positive superhump CVs, reported a likely orbital period of 0.14923 d, finding no signal detection near the period reported by \cite{chenetal01-1}. Time-resolved optical spectroscopy of BB Dor in the high state (Schenker, private communication), although of poor quality, showed single-peaked \Ha~line emission with very broad wings extending up to $\sim \pm 2000$ \kms. This is a well-known characteristic of moderate to low inclination SW Sex stars \citep[e.g.][]{casaresetal96-1,rodriguez-giletal01-1,rodriguez-giletal07-2}. Finally, modelling of FUSE spectra of BB Dor taken long before the decline to the low state yielded an upper limit to the WD temperature of 40000\,K and a distance to the system of $\sim 650$\,pc \citep{godonetal08-1}.

\section{Long term light curve}

We have an ongoing photometric monitoring programme of a group of nova-like CVs which uses the 0.82-m IAC80 telescope on Tenerife, Spain, the 1.3-m SMARTS telescope on Cerro Tololo, Chile (operated by the SMARTS consortium), and the efforts of a large number of Spanish amateur astronomers. A total of 6 systems in the south and 70 northern systems are being monitored.

\begin{figure}
\includegraphics[width=\columnwidth]{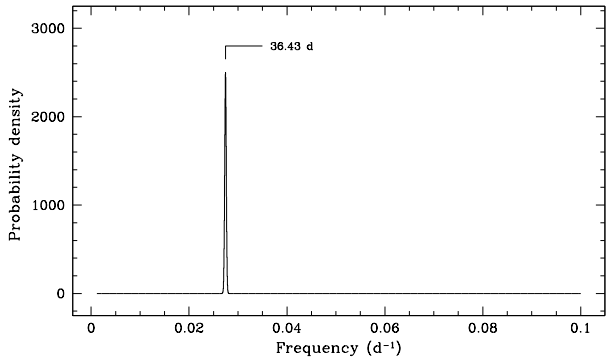}
\caption{Bayesian periodogram of 1.7 years of $V$-band photometric measurements taken before the onset of the first observed decline to the low state.}
\label{fig_bayesian}
\end{figure}

Figure~\ref{fig_longtermphot} presents 5.43 years of photometric monitoring of BB Dor in the interval 2006 August--2012 January. A total of 424 $V$-band images were taken with the 2048$\times$2048 pixel ANDICAM CCD camera on the 1.3-m SMARTS telescope. The exposure time was 60 s, but exposures of 120 s were obtained during the faintest state at $V \sim 19.3$. Each observation consists of two individual images, so two photometric data points are obtained every time the target is observed. The SMARTS consortium provided the images reduced with a pipeline based on \texttt{IRAF}\footnote{See http://www.astro.yale.edu/smarts/ANDICAM/data.html for details on the reduction pipeline.}. The $V$-band light curve was calculated relative to the comparison star $\mathrm{C_1}$ (USNO--A2.0 0300-01740905). Conversion of its $B_\mathrm{A2.0}$ and $R_\mathrm{A2.0}$ magnitudes into the Landolt standard system yielded\footnote{See http://www.pas.rochester.edu/$\sim$emamajek/memo$\_$USNO-A2.0.html.} a $V$-band magnitude of 12.9 for $\mathrm{C_1}$. The nearby star USNO--A2.0 0300--01740223 was adopted as the check star ($\mathrm{C_2}$).

BB Dor started its decline towards the low state around 2008 April 13. In the 1.7 years prior to the onset of the low state the system was at an average magnitude of $V \simeq 14.2$, showing quasi-periodic brightenings up to $V \simeq 13.5$ every $\sim 35-40$ days. A Bayesian generalization of the Lomb-Scargle periodogram \citep{bretthorst88-1} from all the photometric points taken before the first decline started was computed and is presented in Fig.~\ref{fig_bayesian}. The probability density shows a peak at 0.02745 d$^{-1}$, corresponding to a period of 36.43 d. Further brightenings are also apparent during the decline phase, and even when the system settled into the low state ($V \sim 19.3$; see the arrows in Fig.~\ref{fig_longtermphot}). Similar quasi-periodic brightenings have been observed in other nova-like CVs in the high state with similar recurrence times \citep[see e.g.][]{honeycuttetal98-2,hoardetal00-2}. The cause that produces this variability is still unknown, but mass transfer oscillations due to migration of starspots under the L1 point \citep{livio+pringle94-1} and {\em stunted} dwarf nova outbursts \citep{honeycuttetal98-2} are among the proposed mechanisms. With our photometric data alone we can not go deeper into the origin of this variability in BB Dor in the high state. Simultaneous photometry and spectroscopy during several brightenings will definitely help uncover the mechanism involved.  

After about one year in the low state BB Doradus started its recovery but got stuck in an `intermediate' state at $V \sim 16.5$. Several re-brightenings and fadings took place during this phase before entering a second low state. At this second quiescence event BB Dor looked more active, exhibiting many excursions between the low and intermediate states as Fig.~\ref{fig_longtermphot} shows. The system seems to be always fighting for accretion. In fact, a later, second attempt at returning back to the high state put the system at another intermediate state at $V \sim 15.5$, about one magnitude brighter than the previous intermediate level. At the time of writing this paper, BB Dor was recovering from a narrower, third low state. Note that the binary gets stuck in gradually brighter intermediate states on its way to the high state. Long-term light curves of other CVs experiencing low states can be found in the literature \citep[see e.g.][]{,hoardetal04-1,honeycutt+kafka04-1} for comparison.
\begin{table}
\setlength{\tabcolsep}{0.95ex}
\caption[]{Log of the time-resolved spectroscopy.}
\label{table_obslog}
\vspace*{-2.5ex}

\begin{tabular}[t]{lcccccccc}
\hline\noalign{\smallskip}
Night &   Grism & \# & Exp. time      &  Coverage \\    
         &                &  of spectra & (s)             &  ($\mathrm{HJD}-2454830$)  \\    
\hline\noalign{\smallskip}
2009 Jan 1   & Gr\#18   & 29   &   600  &   $3.6135-3.8559$  \\
2009 Jan 2   & Gr\#18   & 20   &   900  &   $4.5637-4.8405$  \\
2009 Jan 3   & Gr\#18   &   7   &   900  &   $5.8043-5.8692$  \\
2009 Jan 4   & Gr\#18   & 24   &   900  &   $6.5541-6.6299$  \\
                     &               &         &           &   $6.6811-6.8501$  \\
2009 Jan 5   & Gr\#18   & 15   &   900  &   $7.5925-7.6683$  \\
                     &               &         &           &   $7.8093-7.8725$  \\
2009 Jan 6   & Gr\#18   & 24   &   900  &   $8.5834-8.8629$  \\
2009 Jan 7   & Gr\#18   & 25   &   900  &   $9.5924-9.8653$  \\

\noalign{\smallskip}\hline
\end{tabular}

\smallskip
\end{table}

\begin{figure}
\includegraphics[width=\columnwidth]{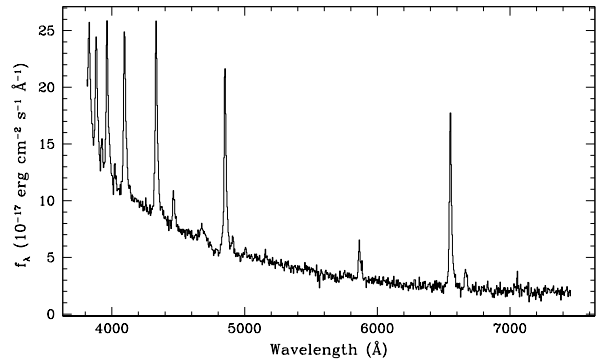}
\caption{Average of the two 900-s spectra of BB Dor taken with grism \#11 at the very beginning of the NTT run. Note the strong Balmer and \ion{He}{i} emission lines with extended wings and the presence of \hel{ii}{4686}+Bowen blend emission, indicating accretion in the system.}
\label{fig_firstspec}
\end{figure}

\section{Spectroscopic observations and their reduction}

\begin{figure*}
\includegraphics[width=8.92cm]{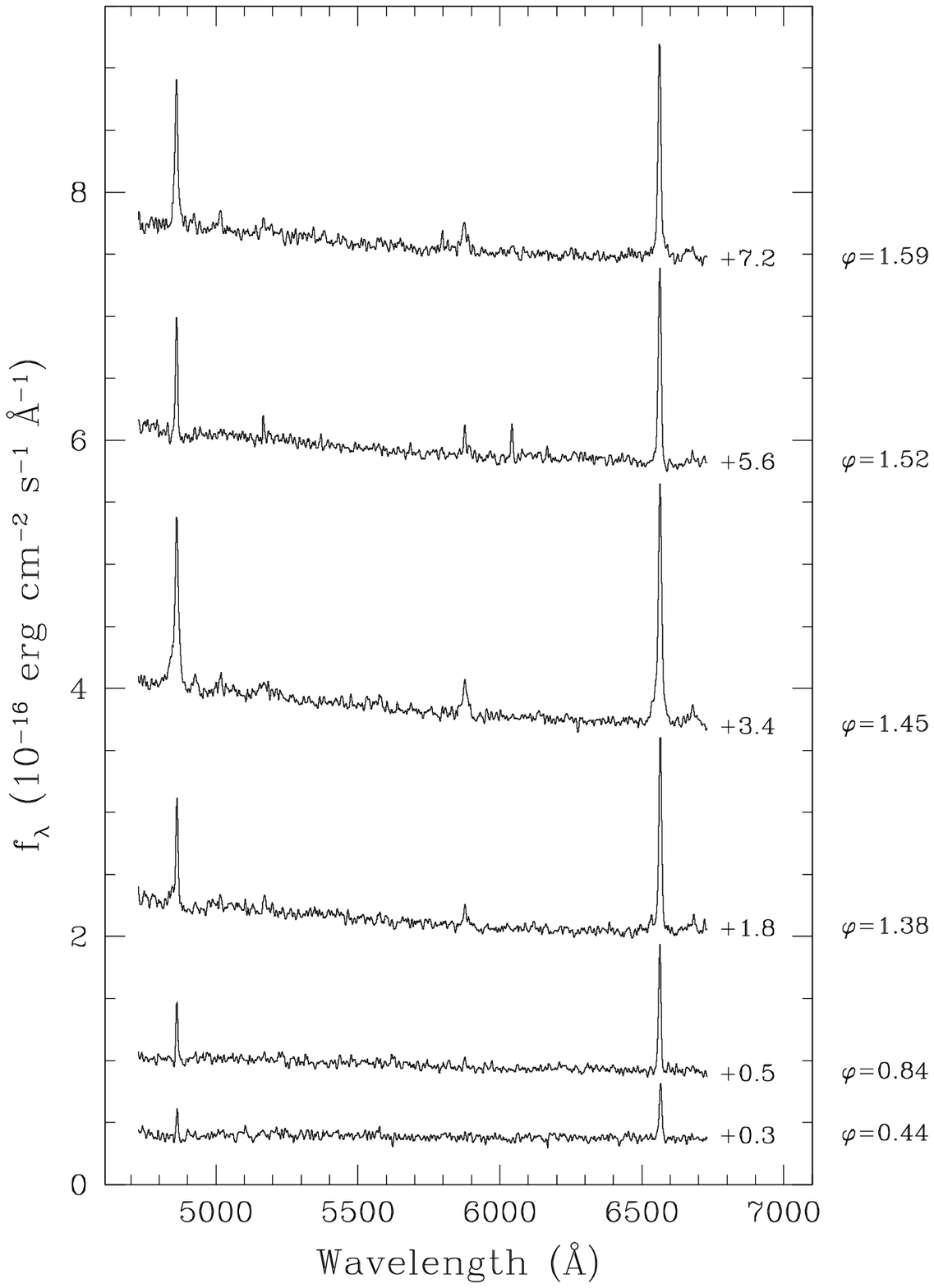}\includegraphics[width=8.92cm]{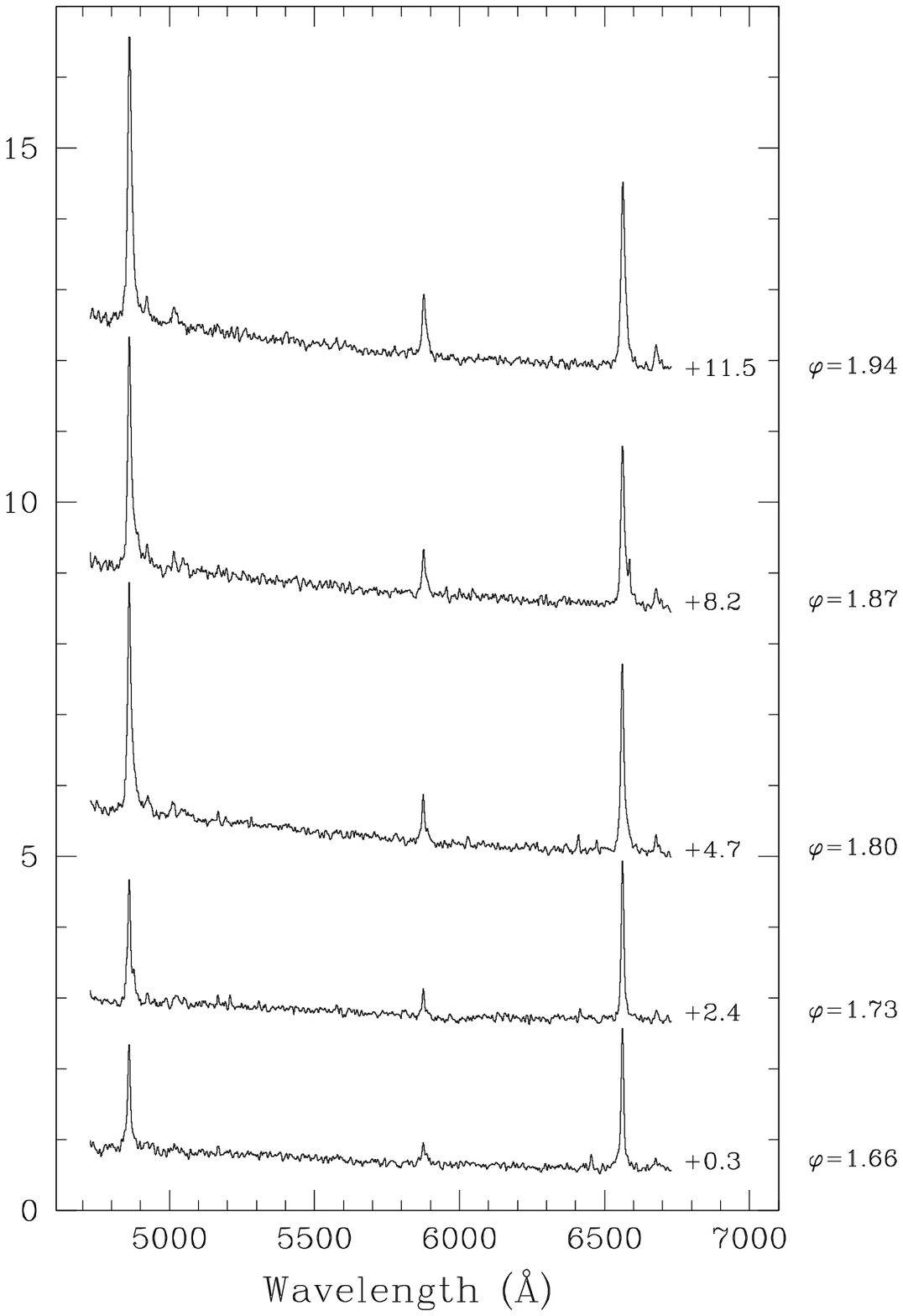}
\caption{Evolution of the accretion event observed during the fourth night of the run. Time runs from bottom to top and from left to right. $\varphi$ is the orbital phase; 0.01 cycle corresponds to 2.2 min. This shows the rapid variability of the lines. The first spectrum at $\varphi=0.44$ shows how the emission lines look like before the onset of the accretion event. All the spectra were shifted in flux by the amount shown to their right.}
\label{fig_blob_n4}
\end{figure*}

We obtained time-resolved spectra with the ESO Faint Object Spectrograph and Camera (EFOSC2) on the 3.58-m New Technology Telescope (NTT) on La Silla. The slit width was fixed at 0.7 arcsec. The log of observations can be found in Table~\ref{table_obslog}.

We processed all the images using standard debiasing and flat-fielding techniques. The one-dimensional spectra were subsequently extracted using conventional optimal extraction techniques in order to optimize the signal-to-noise ratio of the output \citep{horne86-1}. Wavelength calibration was performed in \texttt{MOLLY}\footnote{http://www.warwick.ac.uk/go/trmarsh/software} by means of arc lamp spectra frequently taken to guarantee an accurate wavelength solution. The spectra were then flux calibrated using \texttt{MOLLY}.

\section{The variable spectrum of BB Dor in quiescence}

\subsection{Discovery of intermittent accretion events}

\begin{figure}
\includegraphics[width=\columnwidth]{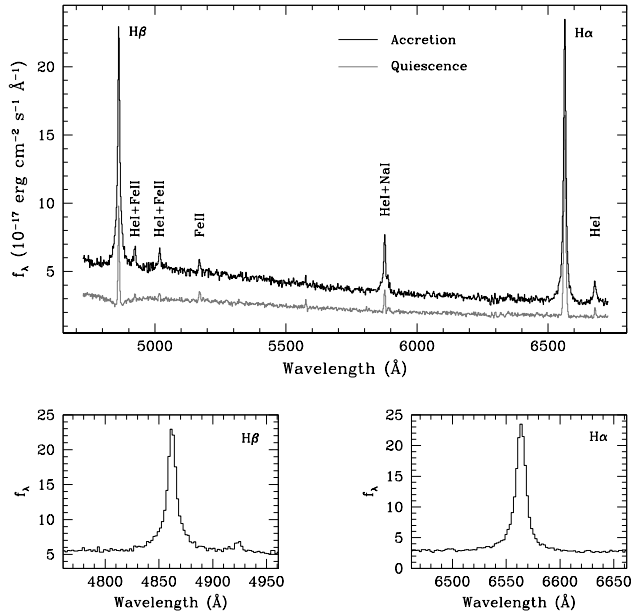}
\caption{Average of all quiescent spectra (gray) compared with the average of all the spectra taken during the accretion events (black).}
\label{fig_low_high_spec}
\end{figure}

The very first spectrum of the 7-day NTT run was taken with grism \#11 ($\lambda\lambda3380-7520$) in order to cover a wide wavelength range. The spectrum of BB Dor showed strong Balmer and \ion{He}{i} emission lines on top of a blue continuum. \hel{ii}{4686} and Bowen blend emission lines were also observed, indicating an excitation level too high for low state conditions. A second spectrum displayed the same features. The average of these two spectra is shown in Fig.~\ref{fig_firstspec}.  We switched immediately to grism \#18 for better spectral resolution. A quick extraction of the first two spectra showed extended wings in the \Ha~and \Hb~emission lines. The intensity of \hel{i}{5876} was also increasing. These strong lines were observed in the next eight 900-s spectra, before starting to weaken in the next two. The remaining spectra until the end of the first night contained much narrower \Ha, \Hb~and weak \hel{i}{5876} lines characteristic of emission from the donor star \citep[see][for a discussion on the narrow emission lines based on higher resolution VLT/FORS spectra]{schmidtobreicketal12-1}, and absent \Line{He}{ii}{4686} and Bowen blend emission. We interpret this behaviour as sporadic accretion events in the system. Remarkably, some of the accretion events observed lasted for only one 900-s spectrum, indicating a fast transition between narrow and broader, stronger emission lines. A total of 11 accretion events were recorded during the whole run (see Table~\ref{table_accretion_events}). Figure~\ref{fig_blob_n4} shows the time evolution of the accretion event that occurred on 2009 Jan 5, while Fig.~\ref{fig_low_high_spec} shows the average of all quiescent spectra compared with the average of all the spectra taken during the recorded accretion events.

Careful inspection of the average spectra presented in Fig.~\ref{fig_low_high_spec} reveals the presence of further narrow emission lines. The strongest of them is the \Line{Fe}{ii}{5169} line, whose presence indicates that the \Line{He}{i}{4922} and \Line{He}{i}{5016} emission lines are likely blended with \Line{Fe}{ii}{4924} and \Line{Fe}{ii}{5018}, respectively. The line at 5577\,\AA~is the remnant of the subtraction of the \ion{O}{i} sky line. In addition, \Line{Na}{i}{5890} emission is observed. We will discuss the origin of the emission lines in BB Dor in section~\ref{sec_emission_origin}. 

\begin{table}
\setlength{\tabcolsep}{0.95ex}
\caption[]{List of accretion events.}
\label{table_accretion_events}
\vspace*{-2.5ex}

\begin{tabular}[t]{lll}
\hline\noalign{\smallskip}
Date &   ~~~~~UT & ~~~~~Notes\\    
\hline\noalign{\smallskip}
\smallskip
2009 Jan 2  & ~~~~~02:12:18$^\mathrm{a}$   &\\
                    & ~~~~~02:42:32--04:43:18                     &\\
                    & ~~~~~08:18:31--08:31:36$^\mathrm{b}$                &\\
2009 Jan 3  & ~~~~~06:59:49                                      & ~~~~~Single spectrum$^\mathrm{c}$\\
2009 Jan 4 & ~~~~~01:51:12                                       & ~~~~~Single spectrum\\
2009 Jan 5 & ~~~~~06:18:37--08:23:18$^\mathrm{b}$                  &\\
2009 Jan 6 & ~~~~~07:24:32--08:42:27$^\mathrm{b}$                  &\\
2009 Jan 7 & ~~~~~05:13:05                                       & ~~~~~Single spectrum\\
                   & ~~~~~06:43:35--06:59:10                     &\\
                   & ~~~~~07:31:11--08:41:40$^\mathrm{b}$                &\\
2009 Jan 8 & ~~~~~07:27:16--08:45:11$^\mathrm{b}$                 &\\                  
\noalign{\smallskip}\hline
\end{tabular}
\smallskip
\begin{minipage}{75mm}
$^\mathrm{a}$\small Average of the two only spectra taken during the accretion event.\\
$^\mathrm{b}$End of the night. \\
$^\mathrm{c}$Single accretion spectrum in between two quiescent ones. Quick accretion event.\\
\end{minipage}

\smallskip
\end{table}
\begin{figure}
\includegraphics[width=6.45cm,angle=-90]{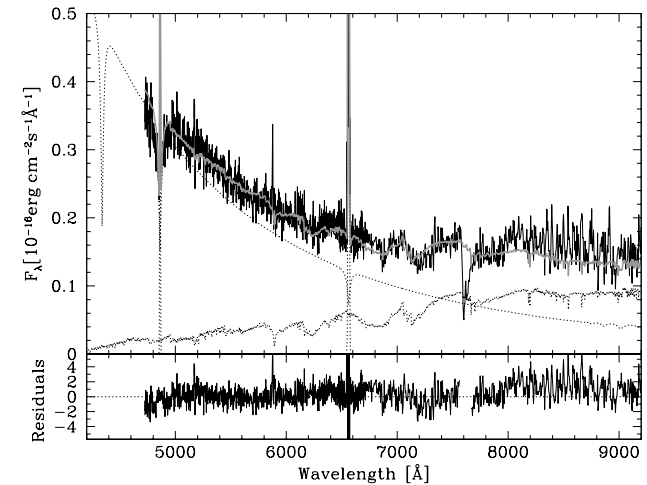}
\caption{Average spectrum of BB Dor in quiescence (solid line) fitted with WD + M-dwarf templates + emission (dotted lines). The gray line is the best composite fit (see text for details).}
\label{fig_specfit}
\end{figure}

\subsection{Spectrum modelling in quiescence}\label{sec_spec_model}

\begin{figure}
\includegraphics[height=8.65cm,angle=-90]{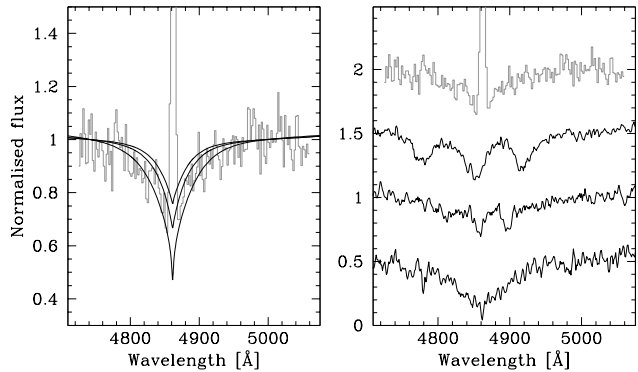}
\caption{{\it Left panel}: Average H$\beta$ profile in quiescence plus superimposed WD spectrum models with effective temperatures of 25000, 35000 and 45000\,K (from bottom to top). {\it Right panel}: Comparison of the BB Dor average H$\beta$ absorption (top) with the H$\beta$ absorption profiles of three magnetic WDs with $B_1$ of 10.8, 5 (showing the narrow absorptions characteristic of Zeeman splitting) and 4.1 MG (from top to bottom; see text for details).}
\label{fig_hbeta}
\end{figure}

Fig.~\ref{fig_specfit} shows the average spectrum of BB Dor in quiescence, that is, when the system is not undergoing an accretion event, constructed from four spectra obtained with grism \#5. The spectra of the WD and the donor star are apparent. The narrow emission lines of the Balmer series and \ion{He}{i} are also visible. In addition, the quiescent spectra taken with grism \#11 ($\lambda\lambda3380-7520$) showed no significant \hel{ii}{4686} or Bowen blend emission. A WD+M-dwarf composite fit resulted in a WD temperature of 30000\,K, a mid-M (likely M3-M4) spectral type for the donor star, and a distance of $1500 \pm 500$ pc. Figure~\ref{fig_hbeta} (left panel) shows the normalised H$\beta$ profile plus non-magnetic models for WD temperatures of (from bottom to top) 25000, 35000 and 45000\,K. The limited signal-to-noise ratio of our NTT blue spectra forces us to give a conservative estimate of $30000 \pm 5000$\,K for the WD temperature, a value in agreement with the results of \cite{godonetal08-1}.

There is evidence that the WDs in some SW Sex stars are magnetic, based on circular polarimetry observations \citep{rodriguez-giletal01-1,rodriguez-giletal02-1,rodriguez-giletal09-1}. However, the low levels of circular polarisation observed suggest WD magnetic field strengths of the order or lower than those observed in the intermediate polar CVs. In order to check for any Zeeman splitting in the quiescent spectra of BB Dor we compare the normalised H$\beta$ absorption profiles of BB Dor (the average of all quiescent spectra) and three magnetic WDs (right panel of Fig.~\ref{fig_hbeta}): SDSS\,J085550.67+824905.1 \citep[$T_\mathrm{eff}=25000$\,K, $B_1=10.8$\,MG,][]{kulebietal09-1}, SDSS\,J080502.28+215320.5 \citep[$T_\mathrm{eff}=28000$\,K, $B_1=5$\,MG,][]{valandinghametal05-1}, and SDSS\, J154305.67+343223.6 \citep[$T_\mathrm{eff}=25000$\,K, $B_1=4.1$\,MG,][]{kulebietal09-1}. We can clearly rule out WD magnetic fields as strong as 10 MG, but could certainly have $B_1 \lesssim 5$\,MG undetected due to the poor signal-to-noise ratio. Better quality blue spectra are needed to check for Zeeman splitting in the low state.

\section{The orbital period of BB Dor}\label{porb}

Because the emission lines show significant changes during the accretion events, we treated the spectra taken during quiescence and accretion separately. Only the spectra taken with grism \#18 were measured for radial velocities. Before measuring the velocities, the spectra were first re-binned into a common velocity scale and then continuum-normalised. Radial velocities of the quiescent \Ha~emission profiles were measured by cross-correlation with a single-Gaussian template with a full-width at half-maximum (FWHM) of 400 \kms. For the radial velocities of the emission line wings in the spectra taken during the accretion events, we cross-correlated with double-Gaussian templates as described in \cite{schneider+young80-2}. In this case the Gaussian FWHM was 200 \kms, while the separation between the two Gaussians was fixed at 1600 \kms.

\begin{figure}
\includegraphics[width=\columnwidth]{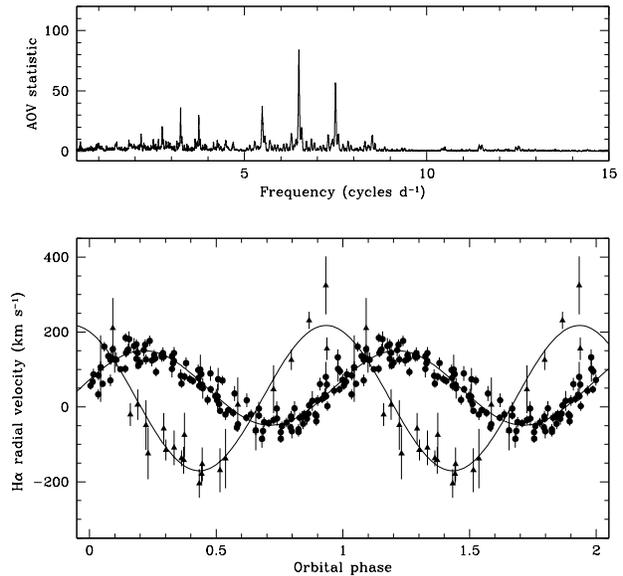}
\caption{{\em Top}: AOV periodogram of the \Ha~radial velocities measured from the quiescent spectra only. {\em Bottom}: \Ha~radial velocities in quiescence (dots) and during the accretion events (triangles) folded on the orbital period. The solid lines are the respective best sine fits (see Table~\ref{table_velfits}). The radial velocity curve of the line wings (triangles) is delayed by 0.18 cycle with respect to the expected motion of the WD, peaking in the blue at orbital phase $\sim 0.45$, which is a defining characteristic of the SW Sex stars in the high state. No phase-binning has been applied. A full cycle has been repeated for clarity.}
\label{fig_aov}
\end{figure}

With the aim of measuring the orbital period of BB Dor we subjected the \Ha~radial velocities obtained for the quiescent spectra to a period analysis using the Schwarzenberg-Czerny (\citeyear{schwarzenberg-czerny96-1}) analysis of variance method implemented in \texttt{MIDAS}. The resulting periodogram is presented in Fig.~\ref{fig_aov}.  The highest peak is found at a frequency of 6.49 d$^{-1}$, which corresponds to a period of approximately 3.70 h. A sine fit to the quiescent radial velocities provided: $P=0.154095 \pm 0.000003$ d ($3.69828 \pm 0.00007$ h) and $T_0(\mathrm{HJD})=2454833.7779 \pm 0.0003 $, where $T_0$ is the time of blue-to-red crossing of the velocities. Our measurement of the orbital period differs from the one reported in \cite{pattersonetal05-3} of 0.14923 d, so their interpretation of the stable photometric signals should be revised. The strongest signal presented in their figure 7 (0.14923 d) may now be associated with a negative superhump, while the other one at 0.16324 d may be a positive superhump. This finding may have consequences on the calibration of the superhump excess ($\varepsilon$)-mass ratio ($q$) relationship for superhumpers, because BB Dor was believed to have the most extreme $\varepsilon$ at 0.094. With our new value of the orbital period the superhump period excess would be revised to $\varepsilon = 0.0593 \pm 0.0005$.   

\begin{table}
\setlength{\tabcolsep}{0.95ex}
\caption[]{Folded radial velocity fits parameters.}
\label{table_velfits}
\vspace*{-2.5ex}

\begin{tabular}[t]{lccc}
\hline\noalign{\smallskip}
 &  $\gamma$ ~~~~& $K$~~~~ & $\varphi_0$~~~~\\
 &  (\kms)  ~~~~      & (\kms) ~~~~ & ~~~~\\    
\hline\noalign{\smallskip}
\smallskip
Quiescence             & $51 \pm 1$ ~~~~& $99 \pm 2$ ~~~~  & --- ~~~~\\
Accretion (wings)    & $24 \pm 8$ ~~~~& $194 \pm 10$ ~~~~ & $0.18 \pm 0.01$ ~~~~\\
\noalign{\smallskip}\hline
\end{tabular}

\smallskip
\end{table}

\subsection{A SW Sex-like S-wave during the accretion events}

As we have already shown, the emission lines get much stronger and their profiles develop extended wings during the accretion events. In fact, we witnessed a change from a quiescent spectrum to an accretion one and back in as quick as fifteen minutes or less (see Table~\ref{table_accretion_events}). During an accretion event the excitation level is increased as well, producing \hel{ii}{4686} and Bowen blend emissions (see Fig.~\ref{fig_firstspec}). As the phasing of the radial velocity curve of the \Ha~emission-line wings indicates (see Fig.~\ref{fig_aov}), the extra emission likely originates on the WD side of the binary system, but its velocity amplitude is almost twice the amplitude observed for the \Ha~emission from the donor star (see Table~\ref{table_velfits}). Remarkably, the wing velocities are delayed by 0.18 cycle with respect to the expected motion of the WD and reach their maximum in the blue at $\varphi \simeq 0.45$. These features are well-known characteristics of the emission S-waves observed both in the SW Sex stars \citep[see e.g.][]{rodriguez-giletal07-1} and the strongly magnetic polar CVs (i.e. AM Her stars) in the high state \citep[e.g. HU Aqr and V2301 Oph,][]{schwopeetal97-1,simicetal98-1}.

\section{The origin of the emission in quiescence}  \label{sec_emission_origin}

\begin{figure}
\includegraphics[width=\columnwidth]{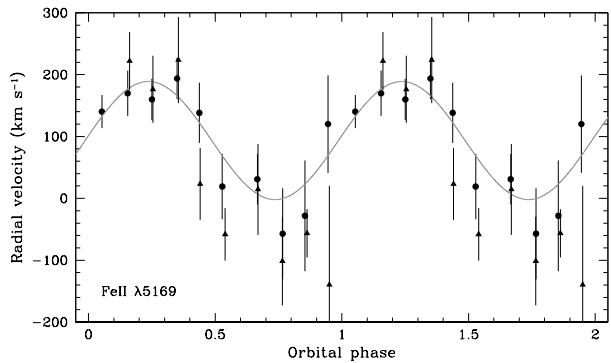}
\caption{Radial velocity curves of the \Line{Fe}{ii}{5169} emission line. Both the quiescent spectra (dots) and the accretion spectra (triangles) were averaged into 10 orbital phase bins to maximise the signal-to-noise ratio of the line profile. The gray curve is the best sine fit to the quiescent velocities only. A full cycle has been repeated for clarity.}
\label{fig_feii_rv}
\end{figure}

The narrow emission lines observed in BB Dor during the low state have a likely origin on the donor star, but whether they form solely via irradiation of the inner face of the donor by the WD or there are other sources of H$\alpha$ emission is unclear. In \cite{schmidtobreicketal12-1} we report the observation of two H$\alpha$ satellite emission lines which clearly detach from the line core. Unfortunately, the spectral resolution is not good enough to resolve them if present in our NTT spectra. Armed with the ephemeris calculated in the previous section we can probe the origin of the emission lines by means of their radial velocity and equivalent width (EW) curves. We will first focus on the \Line{Fe}{ii}{5169} emission line. In order to maximize its signal-to-noise ratio we first binned all the quiescent spectra into 10 phase intervals. Then, we measured the radial velocities of \Line{Fe}{ii}{5169} by cross-correlation with a Gaussian template with $\mathrm{FWHM}=400$~\kms. The resulting radial velocity curve is presented in Fig.~\ref{fig_feii_rv}. The orbital phase of the \Line{Fe}{ii}{5169} radial velocity curve is consistent with that of the quiescent \Ha~line (velocity points marked as triangles in Fig.~\ref{fig_aov}). A very similar radial velocity curve is obtained when the phase-binned \Line{Fe}{ii}{5169} profiles observed during the accretion events only are measured. This indicates an origin on the donor star side of the binary system for the H$\alpha$ and \Line{Fe}{ii}{5169} emission lines both in quiescence and during the accretion events.

\begin{figure}
\includegraphics[width=\columnwidth]{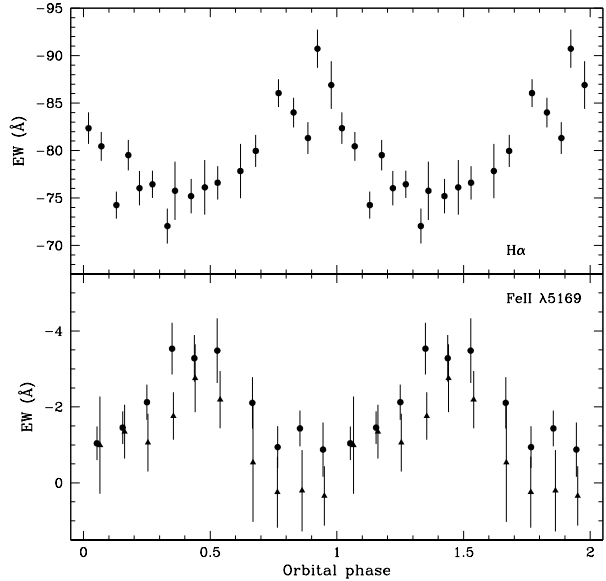}
\caption{{\em Top panel}: EW curve of the \Ha~emission line in quiescence. {\em Bottom panel}: EW curve of the \Line{Fe}{ii}{5169} emission line in quiescence (dots) and during the accretion events (triangles). A full cycle has been repeated for clarity.}
\label{fig_ew}
\end{figure}

If formed on the inner side of the donor star as expected for irradiation, the EW curve should have a maximum close to orbital phase 0.5 and a minimum around 0. To compare the behaviour of the \Ha~and \Line{Fe}{ii}{5169} lines we constructed their EW curves, which are shown in Fig.~\ref{fig_ew}. The EW was measured only for the line cores. Phase binning was applied to both \Line{Fe}{ii}{5169} (10 bins) and \Ha~(20 bins). Unlike \Ha, the \Line{Fe}{ii}{5169} EW curve clearly shows a maximum at phase 0.5. In addition, the line almost vanishes at phase 0. The EW curve obtained from the accretion spectra is very similar as can be seen in Fig.~\ref{fig_ew}. This also supports the adopted $T_0$ in Sect.~\ref{porb}, that is, the narrow emission lines come from the secondary star. This behaviour clearly points to an irradiation origin for the \Line{Fe}{ii}{5169} emission. On the other hand, the \Ha~line remains strong for the whole orbit with a minimum EW at phase $\sim 0.4$ and a maximum at $\sim 0.9$. We observed a similar behaviour in the {\em eclipsing} SW Sex star HS\,0220+0603 (Rodr\'\i guez-Gil et al., in preparation) while comparing the near-IR \ion{Ca}{ii} triplet and the H$\alpha$ emission lines. The \ion{Ca}{ii} lines almost completely vanish at phase zero, with a maximum half an orbit later as expected for an irradiation origin. On the contrary, the H$\alpha$ emission line was visible along the whole orbit with a K--velocity consistent with an origin at the inner Lagrangian point but with no radial velocity delay with respect to \ion{Ca}{ii}. The observation of significant H$\alpha$ emission at zero phase in a high inclination system like HS\,0220+0603 (when the irradiated side is hidden from view) suggests the presence of extra H$\alpha$ emission with an origin either in the chromosphere of the donor star or in material above the orbital plane.

\section{Discussion}

\subsection{The origin of the extra emission during the accretion events \label{sec_line_origin}}

We have shown that BB Dor shows sporadic accretion activity during the low state which translates into significant broadening of the narrow, mainly Balmer and \ion{He}{i} emission lines, and an increase of the excitation level which produces \Line{He}{ii}{4686} and Bowen blend emissions and a bluer continuum. The radial velocity curve of the \Ha~extended wings developed during these accretion events (see Fig.~\ref{fig_aov}) was constructed from accretion spectra widely spread in time, indicating that its phasing is stable. This means that the material shed by the donor star produces the high-velocity emission at a fixed structure in the system. In explaining this behaviour we shall first review what is known about flaring activity in the polar CVs in the low state, for the simple reason that an accretion disc is never present in those systems, so any gas leaving the donor star would be magnetically accreted on to the WD.

Flares during a low state were first observed in the ultraviolet ({\it EUVE}) in the polar CV QS Tel \citep{warrenetal93-1}. AM Her itself also shows variable X-ray activity during the low state \citep{demartinoetal98-3}. Similar flaring activity was later detected in the eclipsing polar UZ For with {\it XMM--Newton} \citep{still+mukai01-1}. The emission during the registered 1.1-kilosecond burst is consistent with the flares typical of rapidly rotating M dwarfs. In addition, the electron temperature is also characteristic of coronal activity on late-type stars. However, the detection of the same flare also in the ultraviolet suggests that the event is probably not coronal in origin, with accretion onto the WD a more likely origin. \cite{pandel+cordova02-1} also observed transient events in X-rays and ultraviolet from UZ For in the low state with {\it XMM--Newton}, with the largest one lasting $\sim 900$ s. The cause is believed to be accretion of material onto the main accretion pole of the magnetic WD as the flare started with the eclipse egress of the main accretion region. \citeauthor{pandel+cordova02-1} link the flaring behaviour of UZ For during the low state to an intermittent increase of the mass transfer due to magnetic stellar activity on the donor star. They show that the total mass accreted is consistent with the mass ejected by the stellar flare. Similar behaviour has been observed in other polar CVs such as VV Pup and V393 Pav, attributed to either stellar flares or coronal mass ejections \citep{pandel+cordova05-1}. Finally, AM Her experienced a flare event with a duration of $\sim 4$ kiloseconds recorded in the $0.5-10$ keV energy range by {\it Suzaku} \citep{teradaetal10-1}. Thus, there is observational evidence of accretion of material expelled by the secondary star due to its magnetic activity onto the poles of the WDs in polar CVs in the low state.

The same phenomenology has also been observed in the optical. \cite{lathametal81-1} observed much broader, stronger emission lines during an accretion event of AM Her. \cite{masonetal94-1} also reported accretion events in AM Her that may have lasted a few minutes only, as observed in BB Dor. In addition, \cite{gaensickeetal99-1} found a variable level of activity during the 1982--1985 low state of the nova-like CV TT Arietis, with spectra typical of almost zero accretion on one night, revealing the WD and the secondary star, and spectra with strong emission lines on another night. Unfortunately, the data were not good enough to look at that variability on shorter time scales. All of this indicates that this behaviour is probably common to low states in both polar and nova-like CVs.

To summarise, stellar flares can provide material for accretion in AM Her stars in the low state. Similar flaring activity in BB Dor prompts us to suggest that this behaviour is caused by material coming from the donor star which is later accreted. As already mentioned, \cite{schmidtobreicketal12-1} discovered two H$\alpha$ satellite emission lines in BB Dor in the low state which we link to the magnetic field of the secondary star, with a likely origin in {\em fixed} magnetic structures such as solar-like prominences concentrated on the L1 point. In this regard, a prominence on the secondary star close to the L1 point has also been invoked to explain the Doppler tomograms of the \Lines{N}{v}{1239,1243} UV emission doublet of AM Her itself in the high state \citep{gaensickeetal98-2}. Such prominences have been reported before in the dwarf novae SS Cygni and IP Pegasi in outburst \citep{steeghsetal96-1}.

\subsection{How is the gas pumped towards the white dwarf during the low state flares?}

\cite{kafkaetal05-1} detected flaring events in the $V$-band light curves taken during the 2004 low state of AM Her concentrated at orbital phases 0.25/0.75, that coincide with the radial velocity maxima of satellite emission lines observed in \Ha~\citep{kafkaetal05-1}. The \Ha~trailed spectra diagrams presented in \cite{kafkaetal08-1} clearly show the satellite emissions with a K-velocity almost tripling that of the \Ha~core. Under the assumption that the two satellite emissions do not cross each other \citeauthor{kafkaetal08-1} argue that they may originate in large coronal loops on the donor star. As mentioned, we \citep{schmidtobreicketal12-1} have detected and resolved the same \Ha~satellite emissions in BB Dor in the low state. It's the first detection of such satellites in an SW Sex system in the low state. Our VLT ToO time-resolved spectroscopy revealed two {\em crossing} narrow satellite emissions having very similar velocity amplitude but different phasing, with their respective maximum excursions to the blue happening at orbital phases $\sim 0.60$ and $\sim 0.85$. A similar phasing can be seen in the \Ha~trailed spectra diagrams presented in figure 4 of \cite{kafkaetal08-1}, suggesting a common origin.
  
\cite{kafkaetal08-1} suggested that mass transfer in AM Her during the low state may be controlled by magnetic field lines associated to both the WD and the donor star. Whatever mechanism produces the satellite emission, the observed similarities between AM Her and BB Dor in the low state naturally lead us to suggest a common process. But the WD in BB Dor is not known to be strongly magnetic. We have shown in Sect.~\ref{sec_spec_model} that the WD in BB Dor, if magnetic at all, would have a field strength of $B_1 \lesssim 5$ MG. In explaining the nature of the satellite lines, we are inclined to invoke an origin linked solely to the magnetic activity of the secondary star in agreement with \cite{schmidtobreicketal12-1}, but the presence of a magnetic WD in BB Dor can not be ruled out at the moment and other mechanisms should be investigated.

As already shown, BB Dor exhibits several of the defining characteristics of the SW Sex stars in the high state. The typical 0.2-cycle phase delay observed in the \Ha~radial velocity curve of BB Dor in the low state during the sporadic accretion events is quite illustrative. If the WD is moderately magnetic ($B_1 \sim 1-5$ MG), accretion of the material provided by the donor star via prominences  or coronal mass ejections concentrated around the L1 point may proceed along the magnetic field lines. This may explain the presence of high-excitation lines like \Line{He}{ii}{4686} and the bluer continuum during the accretion events. In fact, a strong flare detected with the {\it HST} in the ultraviolet in AM Her during a low state (Saar, Kashyap \& Ringwald \citeyear{saaretal06-1}), which was placed at a region close to the L1 point, followed by an enhancement of the WD continuum observed almost simultaneously, causally linked the stellar activity on the donor star and the magnetic accretion event. Faster, time-resolved spectroscopy and photometry of BB Dor and other nova-like CVs in the low state would help solve whether a magnetic WD is present or not. The spin of the presumably asynchronously rotating WD would modulate the accretion onto the magnetic pole (or poles) during an accretion event, leading to photometric and emission-line flux variability.

However, a similar phase delay and high amplitude of the radial velocities may be produced if the extra emission originates at the hot spot region of a cold, remnant accretion disc (see Sect.~\ref{intro}) after each sporadic impact of secondary star material. Time-resolved spectroscopy of accretion events in {\em eclipsing} systems in the low state can give the answer. This scenario can be ruled out if the extra emission does not disappear when the secondary star eclipses the hot spot region.   

\section{Conclusions}

The main results of this work are summarised here:

1) 5.43 years of SMARTS $V$-band photometric monitoring of the SW Sex star BB Doradus caught the system about to enter a low state phase with three distinct fadings down to $V \sim 19.3$. The two recoveries from minimum never put the system brightness back to the high state level. BB Dor always got stuck in an intermediate state before fading towards a new minimum. The brightness at the intermediate state is slowly increasing with time.

2) Brightenings of $\sim 0.7$ mag repeating on a period of 36.43 days were observed from the start of the photometric monitoring campaign until the onset of the first low state (a $\sim 500$ day interval). We also observe these during the decline phase and even after the onset of the low state. Although they may be related to variations of mass transfer from the donor star, we can not address any conclusion with the photometric data alone. Coordinated photometric and spectroscopic observations during the brightenings can help understand their origin. 

3) The mass transfer from the donor star is not completely quenched in the low state. We discovered sporadic accretion events in the faint state at $V \sim 19.3$ during which the narrow Balmer and \ion{He}{i} emission lines from the donor star get much stronger and develop high velocity wings, and \Line{He}{ii}{4686} and Bowen blend emissions are observed. A total of 11 accretion events were recorded, with the quickest one lasting only $\sim 900$ s. During the accretion episodes the \Ha~emission line shows a SW Sex-like S-wave with the typical $\sim 0.2$-cycle delay with respect to the expected motion of the WD.

4) BB Dor is composed of an accretion-heated WD with an effective temperature of $30000 \pm 5000$\,K (in agreement with the results of \citealt{godonetal08-1}) and a M3--M4 donor star. The distance to the system is estimated to be $1500 \pm 500$\,pc.

5) We measured an accurate orbital period of $0.154095 \pm 0.000003$ d ($3.69828 \pm 0.00007$ h) for the first time from \Ha~radial velocities during quiescence. With this new value of the orbital period BB Dor no longer holds the most extreme superhump period excess.

6) Narrow \Line{Fe}{ii}{5169} line emission originating on the donor star is shown to have an irradiation origin, while irradiation is not the only \Ha~line emission source. Stellar activity on the donor star in the form of H$\alpha$ satellite lines \citep{schmidtobreicketal12-1} and chromospheric emission may play an important role.

7) We attributed the events of enhanced line emission to sporadic accretion. It can be said that the system is always {\it fighting} to recover the high mass transfer rate it had in the high state. One possible scenario is accretion onto a discless, magnetic WD after coupling of material provided by the donor star to the WD magnetosphere. However, our search for Zeeman splitting in the quiescent H$\beta$ absorption line was unsuccessful and places an upper limit to the WD magnetic field strength of $B_1 \lesssim 5$\, MG. The similarities with the strongly magnetic AM Her stars in the low state suggest that the material is likely provided by stellar magnetic activity on the donor. A second scenario requires the impact of material sporadically provided by the secondary star onto a cold, remnant accretion disc at approximately the hot spot location. Both scenarios can explain the radial velocity phase delay of the enhanced line wings produced during the discrete accretion events.

\section*{Acknowledgments}
We thank the anonymous referee for comments that improved this manuscript. The use of Tom Marsh's \texttt{MOLLY} package is gratefully acknowledged. PRG thanks the ESO/Santiago Visiting Scientist Program for the approval of a scientific visit
during which part of this work was completed. Partially funded by the Spanish MICINN under the Consolider-Ingenio 2010 Program grants CSD2006-00070: First Science with the GTC and CSD2009-00038: ASTROMOL. We are indebted to the SMARTS service observers Juan Espinoza, David Gonz\'alez, Manuel Hern\'andez, Rodrigo Hern\'andez, Alberto Miranda, Alberto Pasten, Mauricio Rojas, Jacqueline Ser\'on, Joselino V\'asquez, and Jos\'e Vel\'asquez for obtaining all the 1.3-m SMARTS $V$-band images.
  
\bibliographystyle{mn2e}
\bibliography{/Users/prguez/Library/texmf/bib/mn-jour,/Users/prguez/Library/texmf/bib/aabib}

\begin{thebibliography}{}

\bibitem[\protect\citeauthoryear{{Araujo-Betancor}, {G{\" a}nsicke}, {Long},
  {Beuermann}, {de Martino}, {Sion} \& {Szkody}}{{Araujo-Betancor}
  et~al.}{2005}]{araujo-betancoretal05-2}
{Araujo-Betancor} S.,  {G{\" a}nsicke} B.~T.,  {Long} K.~S.,  {Beuermann} K.,
  {de Martino} D.,  {Sion} E.~M.,    {Szkody} P.,  2005, ApJ, 622, 589

\bibitem[\protect\citeauthoryear{{Baskill}, {Wheatley} \& {Osborne}}{{Baskill}
  et~al.}{2005}]{baskilletal05-1}
{Baskill} D.~S.,  {Wheatley} P.~J.,    {Osborne} J.~P.,  2005, MNRAS, 357, 626

\bibitem[\protect\citeauthoryear{{Bretthorst}}{{Bretthorst}}{1988}]{bretthorst%
88-1}
{Bretthorst} G.~L.,  1988, Bayesian spectrum analysis and parameter estimation.
Vol.~48 of Lecture Notes in Statistics, Springer-Verlag

\bibitem[\protect\citeauthoryear{{Casares}, {Mart\'\i nez-Pais}, {Marsh},
  {Charles} \& {L\'azaro}}{{Casares} et~al.}{1996}]{casaresetal96-1}
{Casares} J.,  {Mart\'\i nez-Pais} I.~G.,  {Marsh} T.~R.,  {Charles} P.~A.,
  {L\'azaro} C.,  1996, MNRAS, 278, 219

\bibitem[\protect\citeauthoryear{{Chen}, {O'Donoghue}, {Stobie}, {Kilkenny} \&
  {Warner}}{{Chen} et~al.}{2001}]{chenetal01-1}
{Chen} A.,  {O'Donoghue} D.,  {Stobie} R.~S.,  {Kilkenny} D.,    {Warner} B.,
  2001, MNRAS, 325, 89

\bibitem[\protect\citeauthoryear{{de Martino}, {G\"ansicke}, {Matt}, {Mouchet},
  {Belloni}, {Beuermann}, {Bonnet-Bidaud}, {Mattei}, {Chiappetti} \&
  {Done}}{{de Martino} et~al.}{1998}]{demartinoetal98-3}
{de Martino} D.,  {G\"ansicke} B.~T.,  {Matt} G.,  {Mouchet} M.,  {Belloni} T.,
   {Beuermann} K.,  {Bonnet-Bidaud} J.-M.,  {Mattei} J.,  {Chiappetti} L.,
  {Done} C.,  1998, A\&A, 333, L31

\bibitem[\protect\citeauthoryear{{G\"ansicke}, {Hoard}, {Beuermann}, {Sion} \&
  {Szkody}}{{G\"ansicke} et~al.}{1998}]{gaensickeetal98-2}
{G\"ansicke} B.~T.,  {Hoard} D.~W.,  {Beuermann} K.,  {Sion} E.~M.,    {Szkody}
  P.,  1998, A\&A, 338, 933

\bibitem[\protect\citeauthoryear{{G\"ansicke}, {Sion}, {Beuermann}, {Fabian},
  {Cheng} \& {Krautter}}{{G\"ansicke} et~al.}{1999}]{gaensickeetal99-1}
{G\"ansicke} B.~T.,  {Sion} E.~M.,  {Beuermann} K.,  {Fabian} D.,  {Cheng}
  F.~H.,    {Krautter} J.,  1999, A\&A, 347, 178

\bibitem[\protect\citeauthoryear{{Godon}, {Sion}, {Barrett}, {Szkody} \&
  {Schlegel}}{{Godon} et~al.}{2008}]{godonetal08-1}
{Godon} P.,  {Sion} E.~M.,  {Barrett} P.~E.,  {Szkody} P.,    {Schlegel} E.~M.,
   2008, ApJ, 687, 532

\bibitem[\protect\citeauthoryear{{Hameury} \& {Lasota}}{{Hameury} \&
  {Lasota}}{2002}]{hameury+lasota02-1}
{Hameury} J.~M.,  {Lasota} J.~P.,  2002, A\&A, 394, 231

\bibitem[\protect\citeauthoryear{{Hessman}, {G\"ansicke} \& {Mattei}}{{Hessman}
  et~al.}{2000}]{hessmanetal00-1}
{Hessman} F.~V.,  {G\"ansicke} B.~T.,    {Mattei} J.~A.,  2000, A\&A, 361, 952

\bibitem[\protect\citeauthoryear{{Hoard}, {Linnell}, {Szkody}, {Fried}, {Sion},
  {Hubeny} \& {Wolfe}}{{Hoard} et~al.}{2004}]{hoardetal04-1}
{Hoard} D.~W.,  {Linnell} A.~P.,  {Szkody} P.,  {Fried} R.~E.,  {Sion} E.~M.,
  {Hubeny} I.,    {Wolfe} M.~A.,  2004, ApJ, 604, 346

\bibitem[\protect\citeauthoryear{{Hoard}, {Szkody}, {Honeycutt}, {Robertson},
  {Desai} \& {Hillwig}}{{Hoard} et~al.}{2000}]{hoardetal00-2}
{Hoard} D.~W.,  {Szkody} P.,  {Honeycutt} R.~K.,  {Robertson} J.,  {Desai} V.,
    {Hillwig} T.,  2000, PASP, 112, 1595

\bibitem[\protect\citeauthoryear{{Honeycutt} \& {Kafka}}{{Honeycutt} \&
  {Kafka}}{2004}]{honeycutt+kafka04-1}
{Honeycutt} R.~K.,  {Kafka} S.,  2004, AJ, 128, 1279

\bibitem[\protect\citeauthoryear{{Honeycutt}, {Robertson} \&
  {Turner}}{{Honeycutt} et~al.}{1998}]{honeycuttetal98-2}
{Honeycutt} R.~K.,  {Robertson} J.~W.,    {Turner} G.~W.,  1998, AJ, 115, 2527

\bibitem[\protect\citeauthoryear{{Horne}}{{Horne}}{1986}]{horne86-1}
{Horne} K.,  1986, PASP, 98, 609

\bibitem[\protect\citeauthoryear{{Kafka}, {Honeycutt}, {Howell} \&
  {Harrison}}{{Kafka} et~al.}{2005}]{kafkaetal05-1}
{Kafka} S.,  {Honeycutt} R.~K.,  {Howell} S.~B.,    {Harrison} T.~E.,  2005,
  AJ, 130, 2852

\bibitem[\protect\citeauthoryear{{Kafka}, {Ribeiro}, {Baptista}, {Honeycutt} \&
  {Robertson}}{{Kafka} et~al.}{2008}]{kafkaetal08-1}
{Kafka} S.,  {Ribeiro} T.,  {Baptista} R.,  {Honeycutt} R.~K.,    {Robertson}
  J.~W.,  2008, ApJ, 688, 1302

\bibitem[\protect\citeauthoryear{{K{\"u}lebi}, {Jordan}, {Euchner},
  {G{\"a}nsicke} \& {Hirsch}}{{K{\"u}lebi} et~al.}{2009}]{kulebietal09-1}
{K{\"u}lebi} B.,  {Jordan} S.,  {Euchner} F.,  {G{\"a}nsicke} B.~T.,
  {Hirsch} H.,  2009, A\&A, 506, 1341

\bibitem[\protect\citeauthoryear{{Latham}, {Liebert} \& {Steiner}}{{Latham}
  et~al.}{1981}]{lathametal81-1}
{Latham} D.~W.,  {Liebert} J.,    {Steiner} J.~E.,  1981, ApJ, 246, 919

\bibitem[\protect\citeauthoryear{{Leach}, {Hessman}, {King}, {Stehle} \&
  {Mattei}}{{Leach} et~al.}{1999}]{leachetal99-1}
{Leach} R.,  {Hessman} F.~V.,  {King} A.~R.,  {Stehle} R.,    {Mattei} J.,
  1999, MNRAS, 305, 225

\bibitem[\protect\citeauthoryear{{Linnell}, {Szkody}, {G{\"a}nsicke}, {Long},
  {Sion}, {Hoard} \& {Hubeny}}{{Linnell} et~al.}{2005}]{linnelletal05-1}
{Linnell} A.~P.,  {Szkody} P.,  {G{\"a}nsicke} B.,  {Long} K.~S.,  {Sion}
  E.~M.,  {Hoard} D.~W.,    {Hubeny} I.,  2005, ApJ, 624, 923

\bibitem[\protect\citeauthoryear{{Livio} \& {Pringle}}{{Livio} \&
  {Pringle}}{1994}]{livio+pringle94-1}
{Livio} M.,  {Pringle} J.~E.,  1994, ApJ, 427, 956

\bibitem[\protect\citeauthoryear{{Mason}, {Andronov}, {Borisov} \&
  {Chanmugam}}{{Mason} et~al.}{1994}]{masonetal94-1}
{Mason} P.~A.,  {Andronov} I.~L.,  {Borisov} N.~V.,    {Chanmugam} G.,  1994,
  in {A.~W.~Shafter} ed., ASP Conf. Ser. 56: Interacting Binary Stars {An
  Accretion Event Observed during a Low State of AM Herculis}.
p.~346

\bibitem[\protect\citeauthoryear{{Pandel} \& {C{\'o}rdova}}{{Pandel} \&
  {C{\'o}rdova}}{2002}]{pandel+cordova02-1}
{Pandel} D.,  {C{\'o}rdova} F.~A.,  2002, MNRAS, 336, 1049

\bibitem[\protect\citeauthoryear{{Pandel} \& {C{\'o}rdova}}{{Pandel} \&
  {C{\'o}rdova}}{2005}]{pandel+cordova05-1}
{Pandel} D.,  {C{\'o}rdova} F.~A.,  2005, ApJ, 620, 416

\bibitem[\protect\citeauthoryear{{Patterson}, {Kemp}, {Harvey}, {Fried}, {Rea},
  {Monard}, {Cook}, {Skillman}, {Vanmunster}, {Bolt}, {Armstrong}, {McCormick},
  {Krajci}, {Jensen}, {Gunn}, {Butterworth}, {Foote} \& {Bos}}{{Patterson}
  et~al.}{2005}]{pattersonetal05-3}
{Patterson} J.,  {Kemp} J.,  {Harvey} D.~A.,  {Fried} R.~E.,  {Rea} R.,
  {Monard} B.,  {Cook} L.~M.,  {Skillman} D.~R.,  {Vanmunster} T.,  {Bolt} G.,
  {Armstrong} E.,  {McCormick} J.,  {Krajci} T.,  {Jensen} L.,  {Gunn} J.,
  {Butterworth} N.,  {Foote} J.,    {Bos} M.,  2005, PASP, 117, 1204

\bibitem[\protect\citeauthoryear{{Rappaport}, {Joss} \& {Verbunt}}{{Rappaport}
  et~al.}{1983}]{rappaportetal83-1}
{Rappaport} S.,  {Joss} P.~C.,    {Verbunt} F.,  1983, ApJ, 275, 713

\bibitem[\protect\citeauthoryear{{Rodr{\'{\i}}guez-Gil}, {Casares},
  {Mart{\'\i}nez-Pais}, {Hakala} \& {Steeghs}}{{Rodr{\'{\i}}guez-Gil}
  et~al.}{2001}]{rodriguez-giletal01-1}
{Rodr{\'{\i}}guez-Gil} P.,  {Casares} J.,  {Mart{\'\i}nez-Pais} I.~G.,
  {Hakala} P.,    {Steeghs} D.,  2001, ApJ, 548, L49

\bibitem[\protect\citeauthoryear{{Rodr{\'{\i}}guez-Gil}, {Casares},
  {Mart{\'{\i}}nez-Pais} \& {Hakala}}{{Rodr{\'{\i}}guez-Gil}
  et~al.}{2002}]{rodriguez-giletal02-1}
{Rodr{\'{\i}}guez-Gil} P.,  {Casares} J.,  {Mart{\'{\i}}nez-Pais} I.~G.,
  {Hakala} P.~J.,  2002, in ASP Conf. Ser. 261: The Physics of Cataclysmic
  Variables and Related Objects {Detection of variable circular polarization in
  the SW Sex star V795 Herculis}.
p.~533

\bibitem[\protect\citeauthoryear{{Rodr{\'{\i}}guez-Gil}, {G{\"a}nsicke},
  {Hagen}, {Araujo-Betancor}, {Aungwerojwit}, {Allende Prieto}, {Boyd},
  {Casares} \& {Engels}}{{Rodr{\'{\i}}guez-Gil}
  et~al.}{2007}]{rodriguez-giletal07-1}
{Rodr{\'{\i}}guez-Gil} P.,  {G{\"a}nsicke} B.~T.,  {Hagen} H.-J.,
  {Araujo-Betancor} S.,  {Aungwerojwit} A.,  {Allende Prieto} C.,  {Boyd} D.,
  {Casares} J.,    {Engels} D.,  2007, MNRAS, 377, 1747

\bibitem[\protect\citeauthoryear{{Rodr{\'{\i}}guez-Gil}, {Mart{\'{\i}}nez-Pais}
  \& {de La Cruz Rodr{\'{\i}}guez}}{{Rodr{\'{\i}}guez-Gil}
  et~al.}{2009}]{rodriguez-giletal09-1}
{Rodr{\'{\i}}guez-Gil} P.,  {Mart{\'{\i}}nez-Pais} I.~G.,    {de La Cruz
  Rodr{\'{\i}}guez} J.,  2009, MNRAS, 395, 973

\bibitem[\protect\citeauthoryear{{Rodr{\'{\i}}guez-Gil}, {Schmidtobreick} \&
  {G{\"a}nsicke}}{{Rodr{\'{\i}}guez-Gil} et~al.}{2007}]{rodriguez-giletal07-2}
{Rodr{\'{\i}}guez-Gil} P.,  {Schmidtobreick} L.,    {G{\"a}nsicke} B.~T.,
  2007, MNRAS, 374, 1359

\bibitem[\protect\citeauthoryear{{Saar}, {Kashyap} \& {Ringwald}}{{Saar}
  et~al.}{2006}]{saaretal06-1}
{Saar} S.~H.,  {Kashyap} V.~L.,    {Ringwald} F.~A.,  2006, in IAU Joint
  Discussion Vol.~4 of IAU Joint Discussion, {A Flare-induced Mass
  Transfer/Accretion Event in AM Her?}

\bibitem[\protect\citeauthoryear{{Schmidtobreick}, {Rodr\'\i guez-Gil}, {Long},
  {G\"ansicke}, {Tappert} \& {Torres}}{{Schmidtobreick}
  et~al.}{2012}]{schmidtobreicketal12-1}
{Schmidtobreick} L.,  {Rodr\'\i guez-Gil} P.,  {Long} K.~S.,  {G\"ansicke}
  B.~T.,  {Tappert} C.,    {Torres} M. A.~P.,  2012, MNRAS, submitted

\bibitem[\protect\citeauthoryear{{Schneider} \& {Young}}{{Schneider} \&
  {Young}}{1980}]{schneider+young80-2}
{Schneider} D.~P.,  {Young} P.,  1980, ApJ, 238, 946

\bibitem[\protect\citeauthoryear{{Schwarzenberg-Czerny}}{{Schwarzenberg-Czerny%
}}{1996}]{schwarzenberg-czerny96-1}
{Schwarzenberg-Czerny} A.,  1996, ApJ, 460, L107

\bibitem[\protect\citeauthoryear{{Schwope}, {Mantel} \& {Horne}}{{Schwope}
  et~al.}{1997}]{schwopeetal97-1}
{Schwope} A.~D.,  {Mantel} K.~H.,    {Horne} K.,  1997, A\&A, 319, 894

\bibitem[\protect\citeauthoryear{{\v{S}imi\'c}, {Barwig}, {Bobinger}, {Mantel}
  \& {Wolf}}{{\v{S}imi\'c} et~al.}{1998}]{simicetal98-1}
{\v{S}imi\'c} D.,  {Barwig} H.,  {Bobinger} A.,  {Mantel} K.~H.,    {Wolf} S.,
  1998, A\&A, 329, 115

\bibitem[\protect\citeauthoryear{{Steeghs}, {Horne}, {Marsh} \&
  {Donati}}{{Steeghs} et~al.}{1996}]{steeghsetal96-1}
{Steeghs} D.,  {Horne} K.,  {Marsh} T.~R.,    {Donati} J.~F.,  1996, MNRAS,
  281, 626

\bibitem[\protect\citeauthoryear{{Still} \& {Mukai}}{{Still} \&
  {Mukai}}{2001}]{still+mukai01-1}
{Still} M.,  {Mukai} K.,  2001, ApJ, 562, L71

\bibitem[\protect\citeauthoryear{{Stobie}, {Morgan}, {Bhatia}, {Kilkenny} \&
  {O'Donoghue}}{{Stobie} et~al.}{1987}]{stobieetal87-1}
{Stobie} R.~S.,  {Morgan} D.~H.,  {Bhatia} R.~K.,  {Kilkenny} D.,
  {O'Donoghue} D.,  1987, in {A.~G.~D.~Philip, D.~S.~Hayes, \& J.~W.~Liebert}
  ed., IAU Colloq. 95: Second Conference on Faint Blue Stars {The
  Edinburgh-Cape blue object survey}.
pp 493--496

\bibitem[\protect\citeauthoryear{{Terada}, {Ishida}, {Bamba}, {Mukai},
  {Hayashi} \& {Harayama}}{{Terada} et~al.}{2010}]{teradaetal10-1}
{Terada} Y.,  {Ishida} M.,  {Bamba} A.,  {Mukai} K.,  {Hayashi} T.,
  {Harayama} A.,  2010, ApJ, 721, 1908

\bibitem[\protect\citeauthoryear{{Townsley} \& {Bildsten}}{{Townsley} \&
  {Bildsten}}{2003}]{townsley+bildsten03-1}
{Townsley} D.~M.,  {Bildsten} L.,  2003, ApJ, 596, L227

\bibitem[\protect\citeauthoryear{{Townsley} \& {G{\"a}nsicke}}{{Townsley} \&
  {G{\"a}nsicke}}{2009}]{townsley+gaensicke09-1}
{Townsley} D.~M.,  {G{\"a}nsicke} B.~T.,  2009, ApJ, 693, 1007

\bibitem[\protect\citeauthoryear{{Vanlandingham}, {Schmidt}, {Eisenstein},
  {Harris}, {Anderson}, {Hall}, {Liebert}, {Schneider}, {Silvestri}, {Stinson}
  \& {Wolfe}}{{Vanlandingham} et~al.}{2005}]{valandinghametal05-1}
{Vanlandingham} K.~M.,  {Schmidt} G.~D.,  {Eisenstein} D.~J.,  {Harris} H.~C.,
  {Anderson} S.~F.,  {Hall} P.~B.,  {Liebert} J.,  {Schneider} D.~P.,
  {Silvestri} N.~M.,  {Stinson} G.~S.,    {Wolfe} M.~A.,  2005, AJ, 130, 734

\bibitem[\protect\citeauthoryear{{Warren}, {Vallerga}, {Mauche}, {Mukai} \&
  {Siegmund}}{{Warren} et~al.}{1993}]{warrenetal93-1}
{Warren} J.~K.,  {Vallerga} J.~V.,  {Mauche} C.~W.,  {Mukai} K.,    {Siegmund}
  O.~H.~W.,  1993, ApJ, 414, L69

\end{thebibliography}

\label{lastpage}
\end{document}